\documentclass[twocolumn,showpacs,preprintnumbers,amsmath,amssymb]{revtex4}

\usepackage{graphicx}
\usepackage{dcolumn}
\usepackage{bm}
\usepackage[french]{babel}

\begin{document}

\preprint{}

\title{Catalytic reactions with
bulk-mediated excursions: \\Mixing fails to restore chemical equilibrium}

\author{M.Coppey$^1$, O.B\'enichou$^2$, J.Klafter$^3$, M.Moreau$^1$ and G.Oshanin$^1$}

\affiliation{$^1$ Laboratoire de Physique Th{\'e}orique des
Liquides,
Universit{\'e} Paris 6, 4 Place Jussieu, 75252 Paris, France}

\affiliation{ $^2$ Laboratoire de Physique de la Mati{\`e}re
Condens{\'e}e,  Coll{\`e}ge de France, 11 Place M.Berthelot, 75252
Paris Cedex 05, France}

\affiliation{  $^3$ School of Chemistry, Tel Aviv University, Tel
Aviv 69978, Israel}

\date{\today}

\begin{abstract}
In this paper we analyze 
the effect of 
the bulk-mediated
excursions (BME) 
of reactive 
species on the long-time behavior 
of the catalytic Langmuir-Hinshelwood-like
$A + B \to 0$ 
reactions in systems 
in which a catalytic plane (CP) 
is in contact with 
$\it liquid$ phase, containg concentrations
of reactive particles. 
Such BME result 
from 
repeated
particles 
desorption from the CP, 
subsequent diffusion in the liquid
phase and eventual readsorption on 
the CP away from the
intial detachment point. This process 
which leads to an effective superdiffusive
transport
along the CP.
We consider both "batch" reactions, in which all particles 
of reactive species were initially adsorbed onto the CP, 
and reactions followed by a steady inflow
of particles onto the CP. 
We show that 
for "batch" reactions
the BME provide an effective mixing channel 
and here the mean-field-type behavior emerges.
On contrary, 
for reaction
followed by a steady inflow of particles,
we observe essential departures 
from the mean-field behavior
and find that 
the mixing effect of the BME is insufficient
to restore chemical 
equilibrium. 
We show that a steady-state is 
established as $t \to \infty$, 
in which the limiting
 value of the 
mean coverages of the CP
depends on the particles' 
diffusion coefficient in the bulk liquid 
phase and 
the spatial 
distributions of adsorbed particles are strongly 
correlated. 
Moreover, we show that the relaxation to 
such a steady-state is described by a power-law  
function of time, in contrast to the 
exponential time-dependence
describing the approach to equilibrium in perfectly stirred systems.

\end{abstract}

\maketitle

\section{Introduction}

Catalytically-activated reactions
 play an important role in various
processes in chemistry,
physics and biology.
Such reactions
are involved,
as well,
in many industrial and technological processes, in which
the  design of
desired chemicals requires
the binding of
chemically inactive molecules,
which  recombine 
only
when some third substance -
the catalytic substrate - is present
 \cite{clark,bond,Zangwill}.

One of the simplest examples of such catalytically-activated
reactions, which will be discussed here, is provided
by the so-called Langmuir-Hinshelwood
scheme  \cite{clark,bond,Zangwill}.
This reaction
scheme involves two types of reactive species -
an $A$ and a $B$, which are spread in a gaseous phase in contact
with a solid surface - a catalyst, may adsorb onto the surface
at specific adsorption sites (at  constant
rates $Q_{ads}^{(A,B)}$), desorb from them back to the gas phase (at constant
rate $Q^{(A,B)}$),
and enter into the reaction
\begin{equation}
\label{reaction}
A+B \to P,
\end{equation}
at a
finite reaction rate $K$, as soon as
any
two of unlike
adsorbed species appear at
neighboring adsorption sites. The reaction product $P$
desorbs from the surface
instantaneously and leaves the system.

Within the conventional
mean-field approach \cite{clark,bond,Zangwill},
(in which one discards
the correlations in
particles' distributions on the
catalytic surface),
one gets, in particular,
in the simple limit
$Q_{ads}^{(A)} = Q_{ads}^{(B)} = Q_{ads}$,
$Q_{ads}^{(A)} \gg Q^{(A,B)}$ and at
low particles densities in the gaseous phase,
the following large-$t$ asymptotical behavior
\begin{equation}
\label{1}
C_A(t) = C_B(t) \approx \sqrt{\frac{Q_{ads}}{K}}
\left(1 - \exp\Big(-\frac{t}{T}\Big)\right),
\end{equation}
where $C_{A,B}(t)$ denote mean surface coverages by the $A$ and
$B$ species at time $t$, respectively, while $T$ determines the
characteristic time at which the value $C_{\infty} =
\sqrt{Q_{ads}/K}$ is approached. The expression in Eq.(\ref{1})
can be readily generalized for arbitrary values of $Q^{(A,B)}$,
$Q_{ads}^{(A,B)}$ and for arbitrary particles densities in the gas
phase, which will result in a somewhat more complex expressions
for $C_{\infty}$ and for the characteristic relaxation time $T$.
Note, however, that the long-time approach to $C_{\infty}$ will be
still described by an exponential function of time. It is also
important to emphasize that the state approached as $t \to \infty$
is believed to be a true $\it chemical$ $\it equilibrium$ state, in which the
$A$ and $B$ particles distributions on the surface are
$\delta$-correlated and $C_{\infty}$ is independent of the kinetic parameters.

Within the last two decades a considerable progress have been 
made in the theoretical analysis of the 
kinetics of {\it non-catalytic} reactions \cite{OZ,sf,toussain,rednerkang,zumo,burborne,klaf,Ovch86,zhang,katja,lebo,burlat,gso,oshanin96,5a,pgdg}.
Here, 
a remarkable phenomenon of stochastic
segregation has been discovered 
\cite{OZ,sf,toussain,rednerkang,zumo,burborne,klaf,Ovch86,zhang,katja,lebo,burlat,gso,oshanin96,5a,pgdg},
and 
the
effects of
correlations and fluctuations in particles spatial distributions
on the reaction course
have been elucidated, which are 
in a striking contrast
with the conventional mean-field picture 
\cite{rice}. These studies resulted 
in the inception of a novel interdisciplinary 
domain 
on the boundary between conventional 
chemistry and physics -  
the fluctuation-dominated chemical kinetics. 

Following the early works on the fluctuation phenomena in chemical reactions \cite{OZ,sf,toussain,rednerkang,zumo},
 Ziff and collaborators \cite{Ziff,fich}
have questioned the predictions of 
the mean-field approach in
Eq.(\ref{1}) for the $\it catalytic$  Langmuir-Hinshelwood scheme.
Focussing on the specific example of the oxidation process of the
carbon monoxide on platinum surfaces, $CO + O_2\to CO_2 + O$, Ziff
et al have observed a behavior which is by far richer and
goes far beyond than the traditional mean-field predictions.
In particular, they have discovered that
 as the $\rm CO$ gas pressure is lowered
the system undergoes a first-order transition from a $\rm CO$  saturated
inactive phase (zero rate of $\rm CO_2$ production) into a reactive steady
state (non-zero rate of $\rm CO_2$ production) followed by a continuous
transition into an $\rm O_2$-saturated inactive phase. This continuous
transition was shown to belong to the same universality class as the directed
percolation and the Reggeon field theory \cite{universality_class}.
Different aspects of the kinetic and equilibrium
behavior in this model have been scrutinized, 
revealing the importance of many-particle effects.
\cite{mea,sad,alb,red,alb2,ben,con,eva,krap,park,dick,frach,sholl,mon,dic,arg,osh1,osh2,pop}.

An essential ingredient of the Langmuir-Hinshelwood model is
that the phase confronting the catalytic surface and acting as a
reservoir of particles is $\it gaseous$. This is not the case,
however, in many instances. For many important applications,
especially in biological and chemical systems, the catalytic
substrate appears to be in contact with a $\it liquid$ phase which
comprises concentrations of reactive species (see, Fig.1).

In such systems the reaction kinetics and equilibrium properties
may be affected by yet another important process, not included in
the previous models; namely, here the particles can performs
long-range concerted excursions inside the bulk liquid phase. 
That is, , as
depicted in Fig.1, the particles, adsorbed onto the
surface being in contact with the bulk liquid phase, can desorb,
diffuse rapidly (with the diffusion coefficient being several
orders of magnitude larger than that for a surface diffusion)
within the bulk phase, and then return to the surface at a new
position far away from the detachment point.

Indeed, it has been shown both experimentally and theoretically
(see, e.g., Refs.\cite{bychuk1,bychuk2,levitz,kimmich,kimmich1}
and references therein) that such bulk mediated excursions
(hereafter abbreviated as the BME) can be the principal kind of
motion for particles.
 In biological systems, such
bulk-mediated excursions might play a significant role since here
most of the systems include membranes surrounded by fluid environments \cite{subsuper}; as an example, one may
consider the receptor-ligand reactions which take place on a
membrane surface \cite{science,cell}, or the catalytic efficiency
of proteins, which cut the DNA molecules (one
dimensional substrates) at specific sequences \cite{brockmann02}.
More generally, BME may be involved in certain "searching"
processes \cite{stanley}.

Several prominent features distinguish the situations with a
liquid and with a gas phase in
 contact with a solid
substrate. First, in the case when the solid surface is in contact
with a liquid, the desorption of the adsorbed particles is
generally much more pronounced than in  the latter case; hence,
one expects that the BME process will be more frequent here.
Second, appearing in the liquid phase after the desorption event,
a desorbed particle will move diffusively, being multiply
scattered by the solvent molecules. In view of the geometry of the
system, here the motion relative to the surface is effectively
one-dimensional, such that, after desorption and excursions in the
bulk, any particle will be $\it certain$ to return back to the
surface, in contrast with the situation with the gaseous phase in
which the particle may travel away from the surface almost
indefinitely.

\bigskip

\begin{figure}[ht]
\begin{center}
\includegraphics*[scale=0.6]{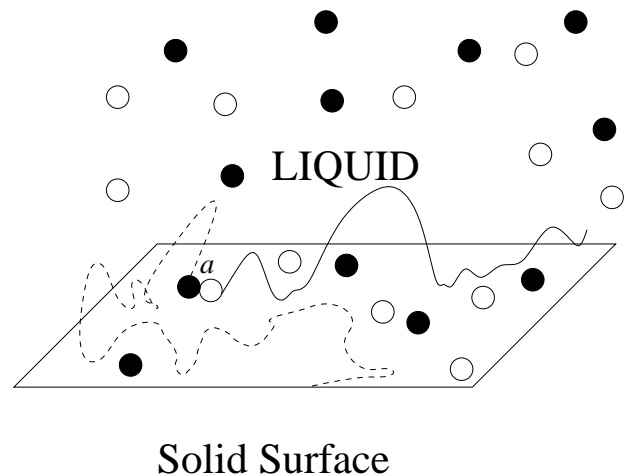}
\caption{\label{Fig1} {\small A sketch of the reactive
Langmuir-Hinshelwood-like
system with bulk-mediated excursions.
Black and white circles denote $A$ and $B$ particles, respectively.
(a) denotes a particle
configuration, in which reaction takes place. Solid and dashed
curved lines depict effective
trajectories of $A$ and $B$ particles, respectively.
}}
\end{center}
\end{figure}

\bigskip

Last but not least, repeated many times, the adsorption/desorption
events separated by the bulk-mediated excursions, will result in
effective motion of any given particle along the surface. A most
striking point here is that this motion is $\it superdiffusive$,
such that with respect to its surface displacements, any given
particle performs a "L{\'e}vy walk" (see Refs.\cite{yossi1,yossi2} for ample physical
discussion). Consequently, instead of a familiar Gaussian
propagator, one finds
\cite{bychuk1,bychuk2,levitz,kimmich,kimmich1}  that here the
distribution $P(r,t)$ of particles' displacements along the
surface is that of a two-dimensional Cauchy process and is
characterized by a long $1/r^3$ tail \cite{55}:
\begin{equation}
P(r,t) = \frac{1}{2 \pi} \frac{ c t}{\left(\Big(c t\Big)^2 + r^2\right)^{3/2}},
\end{equation}
where
$c = D/h$, $D$ being the particles' diffusion coefficient in the bulk liquid,
while $h = b Q_{ads}/Q$ stands for the "adsorption depth" and $b$ is
the "capture range" -  the distance over which a particle
 can directly be adsorbed in a
single displacement step
\cite{bychuk1,bychuk2,levitz,kimmich,kimmich1}. This implies, in
turn, that due to the BME, the distribution $\phi(r)$ of
particles' displacements $r$ along the surface obeys
\cite{bychuk1,bychuk2,levitz,kimmich,kimmich1}:
\begin{equation}
\label{disp} \phi(r) = \frac{r^{\star}}{|r|^3}, \;\;\; r
> r^{\star} = \sqrt{D
t^{\star}},
\end{equation}
$t^{\star}$ being the typical time between the readsorption
events, $t^{\star} = D/(Q_{ads} b)^2$  \cite{bychuk1,bychuk2}. The
impact of such a peculiar transport on the kinetics and
equilibrium properties of the reaction process in
Eq.(\ref{reaction}) has not been elucidated theoretically up to
the present time.

In this paper we analyze the effect of the bulk-mediated
excursions of the reactive species on the long-time behavior 
of the catalytic Langmuir-Hinshelwood-like
reactions, Eq.(\ref{reaction}), in terms of a simplified model,
the model captures only some basic features of the physical system,
but still allows to draw several important conclusions. First of
all, we concentrate here on a totally symmetric situation, in
which the mean concentrations of the reactive particles in the
liquid phase, as well as their adsorption and desorption rates,
are equal to each other. Further on, in this model, we focus on
the events taking place on the catalytic surface and introduce the
presence of a semi-infinite liquid phase containing concentrations
of reactive species in an indirect fashion. That is, we assume
that continuous inflow of $A$ and $B$ particles, dispersed in the
semi-infinite liquid phase, to the catalytic surface can be
modelled as a source, which creates $A$ and $B$ directly on the
surface, independently of each other and at a constant production
rate $Q_{ads}$. Furthermore, we suppose that the BME can be taken
into account by letting the adsorbed particles to perform random,
long-range hopping motion along the adsorption sites of the
catalytic surface with suitably chosen hopping probabilities,
determined by Eq.(\ref{disp}).  Consequently, we propose here a
two-dimensional model which includes two types of
reactive species, which react upon encounters,  perform long-range
(L{\'e}vy or, more spcifically, Gillis-Weiss \cite{gillis,hughes})
walks on the lattice and are continuously introduced onto the
lattice from a reservoir maintained at a constant chemical
potential.

We hasten to remark that this model serves only as
a first approximation of the real physical system 
and there are several other important processes,
which may influence the kinetic behavior of the reaction process in
Eq.(\ref{reaction}). First of all, an assumption
that the "intensity" of particles creation on the lattice is not
varying in time may be inadequate. As a matter of fact, here, in
view of the effectively one-dimensional geometry and diffusive
transport in the liquid phase, non-homogenous particle density
profiles in the direction perpendicular to the catalytic plane,
characterized by a "depletion" zone, may emerge. This will result
in the "intensity" decaying in time. On the other hand, here we
overestimate mixing effect of long-range BME, supposing that for
any particle a jump on distance $r$ along the lattice, once chosen
with the probability distribution in Eq.(\ref{disp}), is executed
$\it instantaneously$, while in reality the transport via BME on
this distance 
takes some time interval, which is a random variable 
having a broad
distribution. Consequent analysis of these effects requires much
more complex approach, which is currently being carried out
\cite{nous}.

Finally, we would like to note that, apart from its relevance to the
reaction process in Eq.(\ref{reaction}) for a catalytic surface in contact with a liquid phase, our analysis sheds the light on
the conceptually  important question of the effect of mixing on
the fluctuation-induced kinetics. As we have already remarked, for
batch $A + B \to 0$ reactions, i.e. reactions in which the
particles of the reactive species are all initially introduced
into the system, in case of equal mean particles densities and
diffusive transport, diffusion appears to be a non-efficient
mixing channel and the like species tend to segregate
spontaneously  in the reaction course, which causes deviations
from the text-book kinetic behavior
\cite{OZ,sf,toussain,rednerkang,zumo,burborne,klaf,Ovch86,zhang,katja,lebo,burlat,gso,oshanin96,5a,pgdg}.
In presence of a steady inflow of reactive species, this effect
gets dramatically increased and the deviations from the
conventionally expected behavior \cite{rice} are getting even more
pronounced \cite{Ovch86,zhang,katja,burlat,gso,oshanin96,arg}. On
the other hand, it has been shown recently in
Refs.\cite{klafter96b,zumo94,klafter97}, which analyzed kinetics
of the batch $A + B \to 0$ reactions, involving particles which
execute L{\'e}vy walks, that in this case, under certain conditions, accelerated diffusion
destroys effectively the particle-particle segregation and the
mean-field behavior prevails. One may now pose quite a legitimate
question whether in situations with a steady inflow of reactive
species the L{\'e}vy walks would provide a sufficiently fast
mixing channel and overcome the strong tendency for segregation of
like species in the course of the process in Eq.(\ref{reaction})?
Our answer is negative. We show that in the situation under study,
despite the fact that we strongly enhanced mixing supposing that
the BME are executed instantaneously, the state reached by the
process in Eq.(\ref{reaction}) as $t \to \infty$ is not a true
chemical equilibrium but only a steady-state. As a matter of fact,
we proceed to show that the long-time concentration $C_{\infty}$ appear to
depend on the kinetic parameters, such as, e.g., particles
diffusivities in the bulk, particles distributions on the lattice
are very strongly (algebraically) correlated and moreover, the
long-time approach to such a steady-state is essentially delayed,
as compared to the exponential dependence in Eq.(\ref{1}) - it is
described by a power-law function of time!

The paper is structured as follows. In section 2 we introduce the
model and basic notations. In section 3, focussing on the case of
batch reactions, we present our analytical approach and reproduce
several known results. In  section 4 we analyze the steady-state
behavior in models with steady particles input, which mimics
Langmuir-Hinshelwood scheme with bulk-mediated excursions, and
discuss the long-time approach to such a steady-state. Finally, we
conclude in section 5 with a brief summary of our results and
discussion.

\section{Model}

Consider a two-dimensional regular lattice which is brought in 
contact with a reservoir
of particles of two types - an $A$ and a $B$, maintained at constant 
chemical potentials $\mu_A$ and $\mu_B$. Here we restrict
our analysis to the special case $\mu_A \equiv \mu_B$. The particles
of both species 
may adsorb onto the lattice at constant rate $Q_{ads}$, desorb 
from the lattice at rate $Q$, an event followed by a long-range
instantaneous
jump of distance $r$ with probability
$\phi(r)$ 
and an immediate re-adsorption. The particles then react at a constant
rate $K$ according to the scheme in Eq.(\ref{reaction}) as soon as
two of them of unlike species appear on the same lattice
site. In most of our analysis we will 
focus on the limit $K \to \infty$, which will allow us
to emphasize the "statistical physics", rather than purely 
"chemical" effects. We will discard 
the 
hard-core 
exclusion between like and unlike species, assuming that the particles'
coverages
are sufficiently small.

Now, 
the long-range jumps performed by the particles of reactive species 
will be described here 
within the framework
of the Gillis-Weiss random walks \cite{gillis} (also referred to
 sometimes as the 
Riemann walks \cite{hughes}), which represent
the lattice version of L{\'e}vy flights \cite{shlesinger} in the limit
$|\mathbf{r}| \gg 1$. We will use here
a bit more general definition of $\phi(r)$, than that
in Eq.(\ref{disp}), and supporse that $\phi(r)$ is given by
\begin{eqnarray}
\label{distt}
\displaystyle
\phi(r)=\frac{\xi}{|\mathbf{r}|^{\mu+d}}.
\end{eqnarray}
Note that the distribution in 
Eq.(\ref{distt}) reduces to the one in 
Eq.(\ref{disp}) in 
the particular case when 
$d = 2$ and
$\mu =1$. In this case, the parameter 
$\xi = r^{\star}$.
Note also that with this definition
of the elementary jump probability, the mean square displacement
per step $\overline{\mathbf{r}^2}$ is infinite for all 
$\mu<2$, which implies that
such a random walk has 
an infinite variance
\cite{shlesinger,montroll}. The long-tailed distribution of the
jump probabilities permits long range jumps and generates a
super-diffusive regime. Gillis-Weiss walks lead to anomalous
diffusion, associated to the dynamic
 exponent
$2/\mu$ for $\mu <2$, 
$\overline{|\mathbf{r}|}^2\sim t^{2/ \mu}$, and to conventional 
diffusion for $\mu \geq 2$, corresponding to
Gaussian random walks, $\overline{\mathbf{r}^2} \sim t$. In the 
case of interest here, i.e. for $\mu = 1$ and $d = 2$, which case
mimics
the reaction in Eq.(\ref{reaction}) mediated by bulk excursions,
one has that in regard to surface displacements, the 
particles  
execute random ballistic-type (with an infinite vellocity) motion with 
$\overline{|r|} \sim t$.

Let now $C_A({\bf r},t)$ and $C_B({\bf r},t)$ denote
the local, (at point with vector ${\bf r}$),
time-dependent coverages of $A$ and $B$ particles, 
respectively. 
Evolution of these
properties 
is governed by the following rate equations
\begin{widetext}
\begin{eqnarray}
\label{dens1}
\displaystyle
\dot{C}_A( \mathbf{r})=-KC_A(\mathbf{r})C_B(\mathbf{r})-\frac{1}{\tau_d}\sum_{\mathbf{r\prime}}\phi(\mathbf{r'}-
\mathbf{r})C_A(\mathbf{r})+\frac{1}{\tau_d}
\sum_{\mathbf{r'}}\phi(\mathbf{r}-\mathbf{r'})C_A(\mathbf{r'}) + Q_{ads}^{(A)}(\mathbf{r},t),
\end{eqnarray}
\begin{eqnarray}
\label{dens2}
\displaystyle
\dot{C}_B(\mathbf{r})=-KC_A(\mathbf{r})C_B(\mathbf{r})-\frac{1}{\tau_d}\sum_{\mathbf{r'}}\phi(\mathbf{r'}-\mathbf{r})C_B(\mathbf{r})+\frac{1}{\tau_d}
\sum_{\mathbf{r'}}\phi(\mathbf{r}-\mathbf{r'})C_B(\mathbf{r'}) + Q_{ads}^{(B)}(\mathbf{r},t),
\end{eqnarray}
\end{widetext}
where the dot denotes the time-derivative, 
the first term on the rhs describes the
decrease in particles' coverages due to the reaction events,
the second and the third terms
describe departures and arrivals of the particles 
at the site ${\bf r}$
at time $t$ due to long-range jumps, respectively.
Note that the summation in the second and the third terms on the rhs of Eqs.(\ref{dens1}) and (\ref{dens2})
extend over all lattice sites, which signifies the long-range character
of particles' migration, while $\tau_d$ denotes the time
each particle typically spends 
on each lattice site between the desorption events, $\tau_d = Q^{-1}$. 
In turn, the fourth terms on the rhs of Eqs.(\ref{dens1}) and (\ref{dens2}) describe 
the (random) 
contributions
to particles' coverages due 
to adsorption of particles from the reservoir, which mimics, in our model,
the presence of particles in the bulk liquid phase. In the usual fashion, 
we admit the following
statistical properties of these "source" terms:
\begin{equation}
\label{p1}
\left \langle Q_{ads}^{(A)}(\mathbf{r},t) \right \rangle = \left \langle Q_{ads}^{(B)}(\mathbf{r},t) \right \rangle = Q_{ads},
\end{equation}
\begin{equation}
\label{p2}
\left \langle Q_{ads}^{(A)}(\mathbf{r},t)  Q_{ads}^{(A)}(\mathbf{r + \lambda},t + \tau) \right \rangle = Q_{ads}^2 + Q_{ads} \delta(\mathbf{\lambda}) \delta(\tau),
\end{equation}
\begin{equation}
\label{p3}
\left \langle Q_{ads}^{(B)}(\mathbf{r},t)  Q_{ads}^{(B)}(\mathbf{r + \lambda},t + \tau) \right \rangle = Q_{ads}^2 + Q_{ads} \delta(\mathbf{\lambda}) \delta(\tau),
\end{equation}
and 
\begin{equation}
\label{p4}
\left \langle Q_{ads}^{(A)}(\mathbf{r},t)  Q_{ads}^{(B)}(\mathbf{r + \lambda},t + \tau) \right \rangle = 0,
\end{equation}
where $\delta(\lambda)$ is the delta-function, 
${\bf \lambda}$ is the correlation parameter and the angle brackets $\langle \ldots \rangle$,
 here and henceforth, denote
the volume averages. Consequently, we stipulate that creation of 
particles on the catalytic surface 
proceeds completely at random, in space and in time, 
at a constant rate, which describe arrival of the particles located 
initially in the bulk liquid phase at progressively
longer distances in the direction perpendicular to the catalytic surface.  

In what follows, we discuss the behavior of the solutions of
the dynamic rate 
equations, Eqs.(\ref{dens1}) and (\ref{dens2}) under  different 
physical conditions.

\section{Batch reactions}

To set up the scene, we discuss first 
the effect of the BME on the kinetics of 
 $A+B\to0$ in the "batch" 
reaction case; namely, in situations in which 
all particles 
of the reactive species which were dispersed  
intially in the bulk liquid phase were 
asorbed onto the surface by some "rapid quench"; 
steady inflow of reactants by the
external source is supposed to be absent here, $Q_{ads}^{(A)}(\mathbf{r},t)
 = Q_{ads}^{(B)}(\mathbf{r},t) \equiv 0$. We suppose, as well, that at 
$t > 0$ particles' desorption from the catalytic surface, and consequently, the BME, are allowed. 

In this case, 
we assume that the initial particles' distributions on the lattice
are random Gaussian, $\delta$-correlated with mean coverages $C_0$; that is,
$C_A(\mathbf{r},0)$ and $C_B(\mathbf{r},0)$ obey:
\begin{equation}
\label{in1}
\Big \langle C_A(\mathbf{r},0) \Big \rangle = \Big \langle C_B(\mathbf{r},0) \Big \rangle = C_{0},
\end{equation}
\begin{equation}
\label{in2}
\Big \langle C_A(\mathbf{r},0) C_A(\mathbf{r + \lambda},0) \Big \rangle = C_{0}^2 + C_{0} \delta(\mathbf{\lambda}),
\end{equation}
\begin{equation}
\label{in3}
\Big \langle C_B(\mathbf{r},0) C_B(\mathbf{r + \lambda},0) \Big \rangle = C_{0}^2 + C_{0} \delta(\mathbf{\lambda}),
\end{equation}
and
\begin{equation}
\label{in4}
\Big \langle C_A(\mathbf{r},0)  C_B(\mathbf{r + \lambda},0) \Big \rangle = 0.
\end{equation}
Now, to analyze the time evolution
 of the mean  
particles' coverages, 
we make use of the analytical approach 
first proposed in Ref.\cite{sf} for the
description of the fluctuation-induced kinetics
of irreversible diffusion-limited $A + B \to 0$
reactions. In this approach the hierarchy
of the reaction-diffusion
equations  
for the higher-order
correlation functions has been truncated at the level of third-order
correlations. Subsequent works (see Refs.\cite{gso,oshanin96,arg}) generalized
the approach to more complex reaction schemes, e.g. to reversible reactions 
or reactions involving interacting particles,
and also showed that such a truncation is tantamount to the assumption
that fields $ C_{A,B}(\mathbf{r},t)$ remain Gaussian at all times; this 
implies that the fourth-order 
correlations 
decouple automatically
into the product of the pairwise correlations which insures that the
third-order correlations vanish.
 
Following Ref.\cite{sf}, we write first the local coverages  
$ C_{A,B}(\mathbf{r},t)$ in the form:
\begin{equation}
\label{fluct}
C_{A,B}(\mathbf{r},t) = C(t) + \delta C_{A,B}(\mathbf{r},t),
\end{equation}
where $\delta C_{A,B}(\mathbf{r},t)$ denote $\it local$ deviations
from the mean coverages $C(t)$. By definition, 
$\Big \langle \delta C_{A,B}(\mathbf{r},t) \Big \rangle \equiv 0$.

Further on, we introduce the pair-correlation functions:
\begin{equation}
\displaystyle
G_{AB}(\mathbf{\lambda},t)=\Big \langle \delta C_A(\mathbf{r},t) \delta C_B(\mathbf{r+\lambda},t) \Big \rangle,
\end{equation}
\begin{equation}
\displaystyle
G_{AA}(\mathbf{\lambda},t)= \Big \langle \delta C_A(\mathbf{r},t) \delta C_A(\mathbf{r+\lambda},t) \Big \rangle,
\end{equation}
and 
\begin{equation}
\displaystyle
G_{BB}(\mathbf{\lambda},t)= \Big \langle \delta C_B(\mathbf{r},t) \delta C_B(\mathbf{r+\lambda},t) \Big \rangle,
\end{equation}
$\lambda$ being the correlation parameter. Note that since we have assumed a totally symmetric situation with regard
to the adsorption/desorption rates, the correlation fucntions $G_{AA}(\mathbf{\lambda},t)$ and 
$G_{BB}(\mathbf{\lambda},t)$ are obviously equal to each other at any time moment,
$G_{AA}(\mathbf{\lambda},t) = G_{BB}(\mathbf{\lambda},t)$.

Next, averaging Eqs.(\ref{dens1}) and (\ref{dens2}),  we obtain
\begin{equation}
\label{dens3}
\displaystyle
\dot{C}(t) =-K\left[C^2(t) + G_{AB}(t)\right],
\end{equation}
where we have used the notation $G_{AB}(t) = G_{AB}(\mathbf{0},t)$, i.e.
$G_{AB}(t)$ is the value of 
the particle-particle
pairwise correlations
at distance $\lambda$ which is equal to the reaction radius $R$ 
(here, in the lattice version of the model, we assumed that
reaction takes place when two particles of unlike species
appear at the same site, i.e.
$R \equiv 0$). 
 
Note now that Eq.(\ref{dens3}) shows that 
the time-evolution of the mean coverage 
is $\it ostensibly$
coupled to the evolution of the 
pairwise correlation function. 
Note also that
if the
correlations are supposed to be insignificant, 
$G_{AB}(t)=0$, as one generally takes 
for the mean-field approach,
one obtains from Eq.(\ref{dens3}) that 
$\dot{C}(t) = - K C^2(t)$, i.e. the conventional
"law of mass action".  This law yields $C(t) \sim 1/K t$
for systems of any spatial dimension
and regardless of the way how the particles move in the system. 
On the other hand, assuming perfect, instantaneous reaction with $K \equiv \infty$,
one gets that
\begin{eqnarray}
\displaystyle
\label{dens4}
C(t)=\sqrt{-G_{AB}(t)},
\end{eqnarray}
which represents the mathematical formulation of the segregation effect;
as a matter of fact, Eq.(\ref{dens4}) shows that the time-evolution 
of the observable - the 
mean coverage, is guided at any time $t$ 
by the time-evolution 
of the pairwise correlations in the system! 
In case when 
both species move diffusively, 
Eq.(\ref{dens4}) 
entails an unusual 
kinetic law $C(t) \sim 1/t^{d/4}$ \cite{OZ,sf,toussain,rednerkang,zumo,lebo},
which thus predicts for $d < 4$ a $\it slower$ time-evolution
of the mean coverage than that 
defined within the mean-field approach. Note also that for finite $K$
equation (\ref{dens4}) holds for times $t$ sufficiently large
such that the fluctuation-induced law
$C(t) \sim 1/t^{d/4}$ determines the long-time asymptotic behavior.

Now, 
to determine
analytically 
the time-evolution of the 
mean coverage in our case with BME, 
we have to
evaluate 
the time dependence of $G_{AB}(t)$ which embodies all 
necessary information on 
the initial fluctuation spectra and particles dynamics.
Making use of Eqs.(\ref{dens1}) to  (\ref{dens3}), 
neglecting 
the 
third-order correlations and employing an evident symmetry
condition 
between the $AA$ and $BB$ correlation functions, ($G_{AA}(\mathbf{\lambda},t) \equiv G_{BB}(\mathbf{\lambda},t)$),  
we find that $G_{AB}(\mathbf{\lambda},t)$ and $G_{AA}(\mathbf{\lambda},t)$
obey the following system of reaction/transport 
equations:
\begin{widetext}
\begin{eqnarray}
\label{k1}
\displaystyle
\dot{G}_{AB}(\mathbf{\lambda},t) = - 2 K C(t)\Big[G_{AB}(\mathbf{\lambda},t)+G_{AA}(\mathbf{\lambda},t)\Big]
+\frac{1}{\tau_d}\sum_{\mathbf{y}}\phi(y) \widehat{{\cal L}_{\lambda}} G_{AB}(\mathbf{\lambda},t),
\end{eqnarray}
and
\begin{eqnarray}
\label{k2}
\displaystyle
\dot{G}_{AA}(\mathbf{\lambda},t) = - 2 K C(t)\Big[G_{AB}(\mathbf{\lambda},t)+G_{AA}(\mathbf{\lambda},t)\Big]
+\frac{1}{\tau_d}\sum_{\mathbf{y}}\phi(y) \widehat{{\cal L}_{\lambda}} G_{AA}(\mathbf{\lambda},t),
\end{eqnarray}
where 
\begin{equation}
\widehat{{\cal L}_{\lambda}} G_{AB}(\mathbf{\lambda},t) = G_{AB}(\mathbf{\lambda+y},t)
+G_{AB}(\mathbf{\lambda-y},t) - 2 G_{AB}(\mathbf{\lambda},t). 
\end{equation}
\end{widetext}
Note that Eqs.(\ref{k1}),(\ref{k2}) together with Eq.(\ref{dens3})
are $\it closed$ with 
respect to the mean coverages
and pairwise correlations, and permits the
evaluation of
these quantities.

In order to solve the system of Eqs.(\ref{k1}),(\ref{k2}) and Eq.(\ref{dens3}), it is expedient
to 
introduce a pair of discrete Fourier transforms:
\begin{eqnarray}
\displaystyle
\nonumber \widetilde{F}(\mathbf{k},t)=\sum_{\mathbf{r}}F(\mathbf{r},t)e^{i(\mathbf{k} \cdot \mathbf{r})}.
\end{eqnarray}
and
\begin{eqnarray}
\displaystyle
\nonumber F(\mathbf{r},t)=\frac{1}{(2\pi)^d}\int_{ \mathcal{B}}\widetilde{F}(\mathbf{k},t)e^{-i (\mathbf{k} \cdot \mathbf{r})}d^dk,
\end{eqnarray}
where $\mathcal{B}$ denotes the first Brillouin zone.

Transforming Eqs.(\ref{k1}) and (\ref{k2}), we get
\begin{eqnarray}
\label{m1}
\displaystyle
\nonumber
\dot{\widetilde{G}}_{AB}(\mathbf{k},t)&=&-2KC(t)\left[\widetilde{G}_{AB}(\mathbf{k},t)+\widetilde{G}_{AA}(\mathbf{k},t)\right]\\
&+&\frac{2}{\tau_d} \Big(  \widetilde{\phi}(\mathbf{k}) - \widetilde{\phi}(\mathbf{0})\Big) \widetilde{G}_{AB}(\mathbf{k},t)
\end{eqnarray}
and
\begin{eqnarray}
\label{m2}
\displaystyle
\nonumber \dot{\widetilde{G}}_{AA}(\mathbf{k},t)&=&-2KC(t)\left[\widetilde{G}_{AB}(\mathbf{k},t)+\widetilde{G}_{AA}(\mathbf{k},t)\right]\\
&+&\frac{2}{\tau_d} \Big(\widetilde{\phi}(\mathbf{k}) - \widetilde{\phi}(\mathbf{0})\Big)\widetilde{G}_{AA}(\mathbf{k},t),
\end{eqnarray}
in which equations we
have made use of the condition $\widetilde{\phi}(\mathbf{-k})=\widetilde{\phi}(\mathbf{k})$,
since random jump process under consideration is symmetric invarient. Equations (\ref{m1}) and (\ref{m2})
are accompanied by the 
initial conditions, which follow from the ones in Eqs.(\ref{in1}) to (\ref{in4}),
\begin{equation}
\label{m3}
\widetilde{G}_{AA}(\mathbf{k},0) = \widetilde{G}_{BB}(\mathbf{k},0) = C_0, \;\;\; \widetilde{G}_{AB}(\mathbf{k},0) = 0.
\end{equation} 
Solution of Eqs.(\ref{m1}) to (\ref{m3}) can be readily obtained explicitly and reads:
\begin{widetext}
\begin{eqnarray}
\label{p}
\displaystyle
\widetilde{G}_{AB}(\mathbf{k},t)=- \frac{C_0}{2} \exp\Big[ - 2 \Big(\widetilde{\phi}(\mathbf{0})-
\widetilde{\phi}(\mathbf{k})\Big) \frac{t}{\tau_d}\Big] \left[1 - \exp\Big( - 4 K \int_0^t C(t') dt'\Big)\right].
\end{eqnarray}
\end{widetext}
We focus first on the perfect reaction case, when $K = \infty$. Here $\widetilde{G}_{AB}(\mathbf{k},t)$ is simply
\begin{eqnarray}
\displaystyle
\widetilde{G}_{AB}(\mathbf{k},t)=-\frac{C_0}{2} \exp\Big[-2 \Big(\widetilde{\phi}(\mathbf{0})-\widetilde{\phi}(\mathbf{k})\Big) \frac{t}{\tau_d}\Big],
\end{eqnarray}
and $G_{AB}(t)$, which enters Eq.(\ref{dens4}) and governs evolution of the mean coverage,
is given by
\begin{eqnarray}
\displaystyle
G_{AB}(t)=\frac{1}{(2\pi)^d}\int_{
\mathcal{B}}\widetilde{G}_{AB}(\mathbf{k},t)d\mathbf{k}.
\end{eqnarray}
It is well-known (see, e.g. Refs.\cite{klafter96b,hughes}) 
that the leading in the small-$k$ limit behavior of 
structure 
function $\widetilde{\phi}(\mathbf{k})=\xi 
\sum_{\mathbf{r}} |\mathbf{r}|^{-\mu-d} \exp(-i \mathbf{r} \cdot \mathbf{k})$
follows:
\begin{eqnarray}
\label{exp}
\displaystyle
\widetilde{\phi}(\mathbf{0})-\widetilde{\phi}(\mathbf{k})  \sim A  |\mathbf{k}|^\mu
\end{eqnarray}
where $A$ is a constant, $A =  \pi \xi/\Gamma(d + \mu) \sin(\pi \mu/2) $, 
$\Gamma(x)$ being the Gamma-function.
Consequently, for $d = 2$ 
the long-time behavior of the correlation function 
$G_{AB}(t)$ obeys:
\begin{eqnarray}
\label{lu}
\displaystyle
\nonumber G_{AB}(t)&=&-\frac{C_0}{2(2\pi)^2}\int_{
\mathcal{B}} \exp\Big[-\frac{2 A t}{\tau_d}|\mathbf{k}|^\mu\Big]d\mathbf{k} \\
\nonumber &\sim&-\frac{C_0}{2(2\pi)}\int_{0}^{\infty}\exp\Big[-\frac{2 A t}{\tau_d}k^\mu\Big]k dk \\
 &=& - \alpha_{\mu} C_0 \left(\frac{\xi t}{\tau_d}\right)^{-2/\mu},
\end{eqnarray}
where $\alpha_{\mu}$ is a dimensionless constant, $\alpha_{\mu} = \Gamma(2/\mu) \Gamma^{2/\mu}(2+\mu) 
\sin^{2/\mu}(\pi \mu/2)/4^{1+1/\mu} \mu \pi^{1+2/\mu}$. Hence,
the long-time decay of the mean coverage $C(t)$ 
is given by
\begin{eqnarray}
\label{genn}
\displaystyle
C(t) \simeq \alpha^{1/2}_{\mu} C_0^{1/2} \left(\frac{\tau_d}{\xi t}\right)^{1/\mu} \sim t^{1/\mu}
\end{eqnarray}
Generalization of the result in Eq.(\ref{genn}) for arbitrary $d$ is straightforward and yields
\begin{eqnarray}
\label{gen}
\displaystyle
C(t)\sim t^{-d/2\mu},
\end{eqnarray}
i.e. precisely the result 
obtained using a different
theoretical approach
 in Refs.\cite{klafter96b,klafter97},
which studied
the kinetics of
batch 
$A + B \to 0$
reactions 
involving particles
performing
L{\'e}vy
walks.
Note also that for $\mu = 2$, which corresponds 
to the case of a standard random walk,
we recover from Eq.(\ref{gen}) the celebrated
fluctuation-induced law 
$C(t) \sim t^{-d/4}$  
\cite{OZ,sf,toussain,rednerkang,zumo,lebo}.
On the other hand, 
for sufficiently small 
values of $\mu$, such as
$\mu<d/2$, fomrula (\ref{genn}) would give a decay faster than the usual
mean field $1/t$ law. Such a behavior is of course specific of the
perfect reaction limit $K=\infty$. Actually the system becomes perfectly
stirred for such small $\mu$, and the segregation phenomenon seizes to
exist, so that the usual $1/t$ decay is recovered. Similar effect 
has been also obtained for
reactions in inhomogeneous systems in Ref.\cite{deem}.

We turn now 
to 
the borderline case $d = 2$ and $\mu =1$,
which is of special
interest here, since it corresponds
to the reactions mediated by bulk excursions. Here, Eq.(\ref{genn})
entails the behavior $C(t) \sim t^{-1}$, i.e. the decay of the 
mean coverge mediated by the Cauchy walks proceeds 
exactly in the same fashion, as
the decay obtained within the conventional 
mean-field approach \cite{rice}. 
This circumstance has prompted the authors of 
Refs.\cite{klafter96b,klafter97}
to conclude that that such a long-range 
transport may serve as an effective
mixing channel which suppresses effectively the "undesired" segregation effect.
In the next section, we will examine
whether this conclusion remains valid in case of reactions followed by a 
steady inflow
of reactive species.

Finally, we note that this special case, in which
the "mean-field", purely chemical 
component, and the decay of correlations 
contribute to the overall kinetics at the same rate,
can be viewed
from a different perspective.
Following the celebrated 
Collins-Kimball treatment
of imperfect diffusion-limited reactions \cite{collins}, 
one may stipulate that 
here 
the time-evolution
of the mean coverage
obeys effectively the second-order 
rate equation of the form
\begin{equation}
\dot{C}(t) = - K_{app} C^2(t),
\end{equation}
where $K_{app}$ is the apparent or effective rate constant, dependent both on $K$
and on the parameters of the Cauchy process.

To determine 
$K_{app}$, we proceed 
as follows: We notice first that for finite $K$ the mean coverage defined by Eq.(\ref{dens3})
cannot decrease faster than the
$t^{-1}$ law expected from the mean-field 
kinetics. This implies that the integral
 $\int_0^tC(t')dt'$ is divergent for $t\to\infty$, and 
hence, at sufficiently large times
\begin{eqnarray}
\exp\Big[ - 4 K \int_0^t C(t')dt'\Big] \ll 1,
\end{eqnarray}
and Eq.(\ref{lu}) describes correctly the 
long-time 
evolution of the pairwise correlation function 
for finite values of $K$. Now,
from Eq.(\ref{lu}) we have that in this case
\begin{equation}
G_{AB}(t) \sim - \frac{C_{0} \tau_d^2}{4 \pi^3 \xi^2} t^{-2}
\end{equation}
Substituting this expression into Eq.(\ref{dens3}) and searching for the solution
of the resulting equation in the form $C(t) = 1/K_{app} t$, we obtain
\begin{equation}
K_{app} = \frac{2 \pi^3 \xi^2}{C_0 \tau^2_d K} \left(\sqrt{1 + 
\frac{C_0 \tau_d^2 K^2}{\pi^3 \xi^2}} -1 \right).
\end{equation}
Recollecting next that $\tau_d = Q^{-1}$ and
$\xi = r^{\star} = \sqrt{D t^{\star}}$, where $t^{\star} = D/Q_{ads}^2 b^2$,
we may rewrite $K_{app}$ 
as
\begin{equation}
K_{app} = \frac{2 \pi^3}{C_0 K} \left(\frac{D Q}{b Q_{ads}}\right)^2 \left(\sqrt{1 
+ \frac{C_0 b^2}{\pi^3} \left(\frac{K Q_{ads}}{D Q}\right)^2} -1\right),
\end{equation}
which relates the kinetic behavior of the "batch" reactions with the BME
to the physical  parameters describing our model.

\section{Reactions with steady inflow of species}

Let us now consider the case when 
the reaction process in Eq.(\ref{reaction}) is accompanied by a 
steady inflow
of particles onto the lattice, which mimics
in our 
model of reactions with bulk-mediated excursions
arrivals onto the surface 
of particles initially dispersed 
in the liquid phase 
from progressively larger and larger distances
in the direction perpendicular to the surface. The statistical 
properties of external sources are defined in section 2, Eqs.(\ref{p1}) to (\ref{p4}).

Averaging Eqs.(\ref{dens1}) and (\ref{dens2}), we find that in this case
the time evolution of the mean coverages is governed by
\begin{eqnarray}
\label{dans}
\displaystyle
\dot{C}(t) = - K \left[C(t)^2+G_{AB}(t)\right]+Q_{ads}. 
\end{eqnarray}
On the other hand, we find that
the 
pairwise correlation function 
$G_{AB}(\mathbf{\lambda},t)$ 
is still determined by Eq.(\ref{k1}), while
$G_{AA}(\mathbf{\lambda},t)$
obeys (see, e.g. Refs.\cite{gso,arg} for the details of the derivation in the diffusion-controlled case):
\begin{widetext}
\begin{eqnarray}
\label{l2}
\displaystyle
\dot{G}_{AA}(\mathbf{\lambda},t) = - 2 K C(t)\Big[G_{AB}(\mathbf{\lambda},t)+G_{AA}(\mathbf{\lambda},t)\Big]
+\frac{1}{\tau_d}\sum_{\mathbf{y}}\phi(y)  \widehat{{\cal L}_{\lambda}} G_{AA}(\mathbf{\lambda},t) + Q_{ads}.
\end{eqnarray}
\end{widetext}
We note, parenthetically, that 
discarding correlations, i.e. setting  in Eq.(\ref{dans})  formally
$G_{AB}(t) \equiv 0$, we recover the 
mean-field
result in Eq.(\ref{1}), which claims that: i) the state approached 
as $t \to \infty$ is a true chemical equilibrium, 
ii) $C_{\infty} = \sqrt{Q_{ads}/K}$ and iii) $C_{\infty}$ is approached exponentially fast.
On the other hand, 
there is an ample evidence
that all three of these statements are wrong 
because of the segregation effects
in case when reactive 
particles perform
conventional diffusive motion on a two-dimensional surface
in presence of a steady
inflow \cite{gso,arg}. 
However, as demonstrated in 
Refs.\cite{klafter96b,klafter97}
and in the  section 3
of the present paper,
for batch reactions with the BME, 
which ultimately 
result 
in random ballistic-like motion along the surface,  
the segregation effect
is suppressed 
and, apart of some renormalization of an the reaction constant $K$,
kinetic behavior follows an essentially mean-field type
dependence $C(t) \sim 1/t$. Consequently, it is not $\it a$ $\it priori$
clear whether $G_{AB}(t)$ governs (or even contributes to) the long-time evolution
of the system in the case $d = 2$ and $\mu = 1$. 

To answer this question, we turn to the time-evolution of  $G_{AB}(\mathbf{\lambda},t)$, defined
by Eqs.(\ref{k1}) and (\ref{l2}). Applying the Fourier transformation, we find, after elementary calculations,
that in the long-time limit (in which the asymptotic behavior in Eq.(\ref{exp}) holds) the Fourier-image of the 
pairwise correlation function $G_{AB}(\mathbf{\lambda},t)$ reads:
\begin{widetext}
\begin{equation}
\label{prop}
\displaystyle
\widetilde{G}_{AB}(\mathbf{k},t) = - \frac{Q_{ads} \tau_d}{4 A |{\bf k}|^{\mu}} \left(1 - \exp\Big[ - 2 A |{\bf k}|^{\mu} \frac{t}{\tau_d}\Big]\right) +
\frac{Q_{ads}}{2} \int^t_0 \exp\Big[ - 2 A |{\bf k}|^{\mu} \frac{t}{\tau_d} - 4 K \int^{t}_{t'} C(t'') dt'' \Big] dt'.
\end{equation}
\end{widetext}
Now, some analysis shows that due to the presence of the function 
$ 4 K \int^{t}_{t'} C(t'') dt''$ in 
the exponential, the second term on the rhs of Eq.(\ref{prop})
is negligibly small compared to the first one. Consequently, we focus on the 
behavior of the dominant contribution.

Consider first the long-time evolution of $G_{AB}(t)$, which enters
the rhs of Eq.(\ref{dans}) and hence, may affect the evolution of the observable - the mean coverage.
Inverting the Fourier transform, we find then 
that in the long-time limit $G_{AB}(t)$ obeys, in $d$-dimensions and for arbitrary $\mu$,
\begin{eqnarray}
\label{last}
\displaystyle
\nonumber &&G_{AB}(t)=-\frac{Q_{ads} \tau_d}{4 A (2\pi)^d}\int_{\mathcal{B}}\frac{d^dk}{|\mathbf{k}|^\mu} \left(1 - 
\exp\Big[ - 2 A |{\bf k}|^{\mu} \frac{t}{\tau_d}\Big]\right)\\ 
&\sim&-\frac{Q_{ads} \tau_d}{8 \pi A}\int_0^{2\pi}\frac{dk}{k^{\mu-d+1}}  \left(1 - 
\exp\Big[ - 2 A |{\bf k}|^{\mu} \frac{t}{\tau_d}\Big]\right)
\end{eqnarray}
Analyzing the behavior of $G_{AB}(t)$, defined in Eq.(\ref{last}),
one notices first that $G_{AB}(t = \infty) = \infty$ for $\mu \geq d$, which signifies
that in such situations $G_{AB}(t)$ is a growing with $t$ function. 
This is precisely the case
discussed in Refs.\cite{gso,arg}, which concerned with the behavior of two-dimensional 
$A + B \to 0$ reactions 
involving diffusive species
in presence of external uncorrelated 
input of particles. 
On the other hand, for $\mu < d$
the stationary value $G_{AB}(t=\infty)$ is finite.

Consequently, in the case of interest here, i.e. for $d = 2$ and $\mu =1$,
we find from Eq.(\ref{last}) the following behavior:
\begin{eqnarray}
\label{j}
\displaystyle
G_{AB}(t) \sim
-\frac{Q_{ads} \tau_d}{4 A}+ \frac{Q_{ads} \tau^2_d}{16\pi A^2 t},
\end{eqnarray}
which predicts that the stationary value of $G_{AB}(t)$, i.e.$
- Q_{ads} \tau_d/4 A$ is approached 
as a power-law, at a rate 
which is proportional to the first inverse power of time!
Note also that $G_{AB}(t) < 0$, which means that pairwise correlation slow down the $\it forward$ reaction.

Substituting the 
expression in Eq.(\ref{j}) into Eq.(\ref{dans}) and solving the resulting equation in the limit $t \to \infty$, 
we find 
eventually that at sufficiently long times the mean coverage obeys
\begin{equation}
C(t) = \sqrt{\frac{Q_{ads}}{K} + \frac{b Q_{ads}^2}{2 \pi D Q}} \left(1 - \frac{t_{char}}{t} + {\cal O}\left(\frac{1}{t^2}\right)\right)
\end{equation}
where we have made use of relations $\xi = r^{\star}$ and $\tau_d = Q^{-1}$,
and the characteristic relaxation time $t_{char}$ is given by
\begin{equation}
t_{char} = \frac{b^4 Q_{ads}^5}{4 \pi^3 Q^2 D^3} \left(\frac{Q_{ads}}{K} + \frac{b Q_{ads}^2}{2 \pi D Q}\right)^{-1}.
\end{equation}
Now, several comments are in order. First of all, we notice that in contrast to the behavior in Eq.(\ref{1}), 
predicted by
the mean-field approach, the actual approach to the $t = \infty$-state is algebraic, proportional 
to the first inverse power of time. This means that the long-time 
relaxation of the mean coverage to its $t \to \infty$ value
is $\it governed$ entirely by the time-evolution of the pairwise correlations.
Second, $C_{\infty}$ is generally 
different of the mean-field value $(= \sqrt{Q_{ads}/K})$ and reduces
to it only when the particles diffusion 
coefficient in the bulk liquid phase $D \to \infty$. The fact that 
$C_{\infty}$ 
depends 
on such a "kinetic" parameter as $D$ is
also quite a prominent feature -  it shows
unambiguously 
that in the reaction process
under study there is no true $\it equilibrium$ but rather
a $steady-state$! 

To emphasize this point,
we turn finally to the analysis of the 
particle-particle 
correlations on the surface in the limit $t = \infty$, embodied
in the pairwise correlation function $G_{AB}({\bf r}) = G_{AB}({\bf r},t = \infty)$.
 From Eq.(\ref{prop}) we find then that these stationary correlations obey
for $d = 2$ and $\mu = 1$:
\begin{equation}
G_{AB}({\bf r}) = - \frac{Q_{ads} \tau_d}{16 \pi^2 A} \int^{2 \pi}_0 d\theta \int^{2 \pi}_0 dk \exp[ i k r \cos(\theta)],
\end{equation}
which yields, in the limit $|{\bf r}| \to \infty$, the following behavior
\begin{equation}
G_{AB}({\bf r}) = - \frac{b Q_{ads}^2}{8 \pi^3 Q D} \frac{1}{|{\bf r}|} + {\cal O}\left(\frac{1}{|{\bf r}|^{3/2}}\right).
\end{equation}
Therefore, the decay of 
correlations in the particles' distributions on the surface
is algebraic, i.e. the particles' distributions show
a quasi-long-range order. As a consequence, 
despite the fact 
that the BME process effectively mixes the system
in the case of batch reactions restoring the mean-field-type
behavior and suppressing the segregation effects,  
in the case with a steady inflow of particles, which mimics the presence of 
particle's concentrations in
the bulk liquid phase, the bulk mediated excursions
fail to establish chemical equilibrium.

\section{Conclusion}

To conclude, in this paper we have analyzed 
the effect of 
the bulk-mediated
excursions of reactive species on the long-time behavior 
of the catalytic Langmuir-Hinshelwood-like
reactions in systems in which a catalytic surface confronts a 
$\it liquid$ phase, containg concentrations
of reactive particles. 
Such bulk mediated excursions result 
from particles desorption from the lattice, subsequent fast diffusion in the liquid
phase and eventual adsorption on the surface far away from the
intial detachment point. Repeated
many times, such BME yield an effectively long-range
particles transport along the catalytic surface
with superdiffusive related properties. We have considered both "batch" reactions, i.e. reactions in which all particles 
of reactive species were initially adsorbed onto the surface, 
and reactions followed by a steady inflow
of particles onto the catalytic surface. 
The latter situations, under certain assumptions, mimic
the presence of
particles concentrations in the bulk liquid phase which act as a reservoir of particles.
We have shown that 
for "batch" reactions, in accord with previous analysis \cite{klafter96b,klafter97},
the BME provide a very effective mixing channel 
which suppresses the segregation effects, such that the mean-field-type behavior prevails.
On contrary, 
for reactions
followed by a steady inflow of particles,
we observe essential departures from the mean-field behavior
and find that 
the mixing effect of the BME is insufficient
to restore chemical 
equilibrium. We show that here a steady-state is 
established as $t \to \infty$, in which the limiting value of 
mean coverages of the catalytic surface
depends on the particles' 
diffusion coefficient in the bulk liquid phase 
and the particles' distributions on the lattice are strongly, algebraically correlated. 
Moreover, the relaxation to 
such a steady-state is described by a power-law 
function of time, in contrast to the exponential time-dependence
describing the approach to equilibrium in perfectly stirred systems.

The authors greatly acknowledge
helpful discussions
with R.Kimmich and P.Levitz. GO wishes to thank 
the Alexander von Humboldt Foundation for financial support 
via the Bessel Research Prize and the
Max-Planck-Institute Stuttgart 
for the warm hospitality.


\begin{thebibliography}{99}
\bibitem{55} We remark
that, of course, for particles adsorbed on a solid surface
effective
 long-range
excursions may result from the conventional surface diffusion -
i.e. random motion of any adsorbed particle along the surface by
activated hopping on the array of the adsorption sites. Here, as
well, any particle may perform activated long-range hops between
the adsorption sites, which at low damping are
 not necessarily constrained
to neighboring sites only. However, in a striking contrast to the
BME, the distribution function of the length of successive jumps
is exponential, such that all moments exist, mean path
length only slightly exceeds the typical distance
 between the adsorption sites and
the overall propagator is a Gaussian \cite{pollak}.
\end{thebibliography}
\end{document}